\documentclass[conference]{IEEEtran}
\IEEEoverridecommandlockouts
\usepackage{cite}
\usepackage{booktabs} 
\usepackage{array}  
\usepackage{url}
\usepackage{amsmath,amssymb,amsfonts}
\usepackage{algorithmic}
\usepackage{graphicx}
\usepackage{textcomp}
\usepackage{xcolor}
\usepackage[linesnumbered,ruled,vlined]{algorithm2e}
\def\BibTeX{{\rm B\kern-.05em{\sc i\kern-.025em b}\kern-.08em
    T\kern-.1667em\lower.7ex\hbox{E}\kern-.125emX}}
\begin{document}

\title{Advancing Obfuscation Strategies to Counter China's Great Firewall: A Technical and Policy Perspective\\
}

\author{\IEEEauthorblockN{Li Li}
\textit{Virginia Tech}\\
}

\maketitle

\begin{abstract}
China's Great Firewall (GFW) exemplifies one of the most extensive and technologically sophisticated internet censorship frameworks worldwide. Serving as a cornerstone of state-directed digital governance, it integrates a multitude of methods - ranging from DNS manipulation and IP blocking to keyword filtering and active surveillance - to control online information flows. These measures, underpinned by both technical proficiency and administrative oversight, form a formidable obstacle to open communication and data privacy. This paper critically examines the GFW's principal detection techniques, including Deep Packet Inspection (DPI), domain name tampering, and traffic fingerprinting, and analyzes how they align with broader governmental mechanisms. In parallel, we evaluate emerging countermeasures that leverage obfuscation, encryption, and routing innovations to circumvent these restrictions. By situating technical strategies within the broader context of governance and human rights, this work underscores the ongoing and evolving contest between state-imposed internet controls and individual efforts to maintain unrestricted access to digital resources.

\end{abstract}


\section{Introduction}
Cybersecurity has profoundly impacted various fields, including blockchain, artificial intelligence, urban applications, and data transmission \cite{6,7,8,9,10}. One of the world's most complex Internet censorship systems, the Great Firewall of China (GFW), forms the cornerstone of the Chinese digital governance strategy. The Chinese cyberspace is tightly controlled because of the GFW, where the GFW utilizes a wide range of strategies and techniques to protect against external incursions. The techniques for enforcing the wall's restrictions include DNS pollution, IP address blocking, keyword filtering, Deep Packet Inspection (DPI), and active probing. These methods allow for real-time monitoring and blocking of content and tools that the censors do not want the public to see. In fact, this increasing complexity of these strategies has made the GFW not only a technological marvel but also a symbol of the challenges posed by state-driven internet surveillance.

In response, VPNs and other privacy tools have become critical for users trying to bypass the GFW. These tools work by doing two essential things. The first is to encrypt all the information that the user sends and receives. This way, even if the GFW does get a look inside the virtual pipe, it won't understand any of the content. But the virtual pipe still has to connect to a server somewhere, and if that server is located in China, it will be in GFW territory. As a result, the second essential thing that a good privacy tool does is to obscure not just the content, but also to obscure where in the virtual space of the Internet the sender and receiver are, so that the GFW has no hope of detecting it \cite{11,12,13}.

\section{VPN Technology and Early Countermeasures}
To understand the dynamic between the Great Firewall of China (GFW) and privacy tools, it is essential to examine the foundational principles of Virtual Private Networks (VPN) and the initial countermeasures employed to bypass early detection techniques.

\subsection{Fundamentals of VPN Technology}

By definition: "A VPN is a communications environment in which access is controlled to permit peer connections only within a defined community of interest, and is constructed through some form of partitioning of a common underlying communications medium, where this underlying communications medium provides services to the network on a nonexclusive basis" \cite {Sisalem2005}. In simple words, virtual Private Networks (VPNs) are tools that users widely use to protect personal data while enhancing online privacy for communication. This process includes two key parts, creating encrypted tunnels that hide user data from unauthorized access \cite{14,15,16}:

\begin{itemize} \item Encryption: VPNs use robust encryption algorithms, such as AES and RSA, to protect data during transmission. This ensures confidentiality and prevents eavesdropping. \item Tunneling Protocols: These protocols, such as OpenVPN and L2TP/IPSec, encapsulate data packets within encrypted tunnels, hiding their contents from network monitoring systems. \end{itemize}

VPNs offer several benefits that are particularly relevant in censored environments like China: \begin{itemize} \item Masking IP Addresses: VPNs route user traffic through servers in different locations, making it appear as if the user is accessing the internet from another region. \item Bypassing Restrictions: Encrypted traffic can often evade simple blocking mechanisms, enabling access to censored content. \end{itemize}

\subsection{Early VPN Countermeasures in China}

The initial strategies of the GFW to detect and block VPNs were relatively straightforward. Techniques included port-based filtering, where traffic on commonly used VPN ports was blocked, and IP address blacklisting, which targeted known VPN server addresses. To counteract these measures, early VPN providers implemented basic obfuscation strategies, such as:

\begin{itemize} \item Port Hopping: Running VPN traffic over ports commonly used by standard web traffic, such as port 443 for HTTPS. \item SSL/TLS Encryption: Making VPN traffic appear indistinguishable from regular encrypted web traffic. \end{itemize}

While these methods were effective in the short term, they proved insufficient against the GFW’s continuous advancements.

\section{Detection Strategies of the Great Firewall of China (GFW)}
This part discusses various  detection mechanisms used by the Great Firewall of China (GFW), a central part of China’s large-scale internet censorship system.
\subsection{DNS Pollution and Hijacking}

The DNS converts human-readable domain names into machine-readable IP addresses. By controlling DNS resolution, the GFW can efficiently restrict access to websites. The GFW uses two methods: DNS pollution and DNS hijacking. 

DNS pollution involves injecting invalid DNS data into the resolver’s cache, which breaks the normal DNS resolution process and can result in receiving false or un-reachable IP addresses when trying to visit blocked domain names. 

DNS hijacking redirects legitimate DNS queries to malicious or unintended servers. Unlike DNS pollution, which mostly impacts resolver caches, DNS hijacking can forever change the direction of DNS traffic. 

\subsection{IP Address and IP Range Blocking}

IP block and IP range block block users from certain network resources by only allowing them to access certain IP addresses or IP ranges. The GFW leverages this method to successfully monitor and control internet traffic. 

IP address blocking restricts or blocks access to individual IP addresses. IP range blocking involves blocking a number of consecutive IP addresses, usually by defining IP address ranges (e.g., subnets). 

\subsection{Keyword Filtering and URL Detection}

Keyword filtering and URL detection are some of the most important technical mechanisms by which the GFW implements internet censorship. The GFW is able to identify and block access to certain content by parsing keywords and URL paths within network requests. 

Keyword filtering works by examining the network traffic for specified sensitive keywords. When these keywords are detected, the system will block or disrupt the related traffic. 

URL filtering refers to the process of sifting through URLs in user requests to prevent access to blocked sites or pages. 

\subsection{Deep Packet Inspection}

Deep Packet Inspection (DPI) is a network packet filter that is able to examine the content of data packets. DPI enables the GFW to discover, classify, and monitor certain network traffic to allow internet content to be censored.

The core of DPI is network data packet analysis, usually in the form of packet capture and parsing, content analysis and recognition, and decision-making and processing.

\subsection{Active Probing and Interference}

Active probing and interference are technologically sophisticated techniques by which the GFW enacts internet censorship. Unlike passive monitoring, active probing and interference involves sending specific data packets or requests in order to find, and disrupt tools and services that bypass censorship mechanisms.

Active probing and interference accomplish this primarily by reading and determining target traffic, sending probe requests, and enabling interference. 

\section{Strategies for Bypassing the Great Firewall: VPN Obfuscation Techniques}
This section discusses strategies about how to get around the GFW with the help of VPN obfuscation techniques. VPN obfuscation disguises VPN traffic, making it less obvious to censorship tools such as the GFW.
\subsection{Protocol Obfuscation}

Protocol obfuscation is the process of obscuring or masking the features of network communication protocols to avoid intercepting, recognizing or censoring traffic. It is primarily intended to oppose the GFW’s network censorship and traffic monitoring by presenting communication as normal, legitimate traffic to evade censorship and blocking. 

This approach can change the structure, pattern, or contents of data packets — it deviates from standard protocol characteristics — and this makes it hard for detection systems to discern particular protocols. This solution effectively defeats protocol-based blocking and interference. 

Additionally, protocol obfuscation can disguise traffic into generic protocols (HTTPS, HTTP) and make it look like ordinary network calls. The disguise technique enables obfuscated traffic to disguise itself among web traffic, greatly reducing the chances of being detected and stopped. 

To increase concealment, protocol obfuscation uses heavy encryption and randomness to mask the true nature and structure of the traffic. This not only increases communication security but also significantly increases the resistance to DPI. 

\subsection{Traffic Obfuscation}

Traffic obfuscation is a technique that processes network traffic to make it difficult to detect using DPI. This involves using third-party obfuscation, which are part of VPN services, to alter the appearance of the traffic so that it resembles common network protocols like HTTPS or HTTP/2. In this way, VPN traffic blends better with internet traffic, reducing the risk of detection.

\subsection{Dynamic and Random Ports}

Through dynamic and random ports, VPN services randomly pick communication ports rather than common VPN ports. VPN traffic can also travel across multiple ports. This strategy of port diversification makes monitoring and blocking even harder. 

\subsection{Traffic Splitting and Multipath Transmission}

By dividing data and routing it through different routes, multi-path VPN ensures that VPN traffic is routed over various network paths and is less likely to be detected by a single monitoring station. Split tunneling, on the other hand, means that you can only send part of the traffic through the VPN tunnel and then connect it directly to the internet. This reduces the overall characteristics of VPN traffic, and thus decreases the detection risk.

\subsection{Anti-Censorship Protocols}

Anti-censorship policies are geared toward censorship dissent. These tools are highly customization and can be customized for multiple obfuscation modes, which makes their traffic difficult to detect and filter. For instance, obfuscation layers may lead to detection difficulties. 

\subsection{Domain Hiding and Dynamic DNS}

With third-party large-scale Content Delivery Networks (CDNs) for domains and infrastructure, domain fronting methods enable VPNs to disguise their actual server addresses so that traffic flows are seen as normal CDN traffic. Dynamic DNS involves frequent updates of server domain names and IP addresses, so it is not easy for static lists-based blocking techniques to prevail. 

\subsection{Using HTTPS and TLS Encryption}

By using high-quality TLS encryption protocols, VPN traffic is not only content-encrypted, but conceals the connection's presence and state behind the encryption layer, allowing DPI to see very little about what type of traffic is being sent. Using newer versions of TLS provides additional security and prevents man-in-the-middle attacks, as well as improving traffic masking. 

\subsection{Frequent Updates and Protocol Switching}

VPN providers use frequent updates and protocol change strategies. They dynamically switch the protocols and ports used by monitoring applications so that monitoring systems do not develop fixed detection and blocking habits. Further, VPN providers leverage automated tools that continuously update obfuscation technology and protocols in response to the latest detection techniques of the censorship system. 

\section{Correspondence Table of GFW Detection Strategies and VPN Obfuscation Techniques}

In this section, we present table, as shown in Table one, that shows detection patterns implemented by the Great Firewall of China (GFW) and various VPN obfuscation algorithms. The table is a summary of this, illustrating the nature of the specific obfuscation protocols designed to thwart the GFW’s surveillance and blocking systems.

\begin{table*}[!htbp]
\caption{Correspondence Table of GFW Detection Strategies and VPN Obfuscation Techniques}
\label{tab:correspondence}
\begin{tabular}{>{\raggedright\arraybackslash}p{4cm} >{\raggedright\arraybackslash}p{6cm} >{\raggedright\arraybackslash}p{6cm}}
\toprule
\textbf{GFW Detection Strategy} & \textbf{Corresponding VPN Obfuscation Technique} & \textbf{Description} \\
\midrule
DNS Pollution and Hijacking & 
Domain Hiding and Dynamic DNS\newline
• Dynamic DNS: Regularly change server domain names and IP addresses to avoid static list blocking. & 
Dynamic DNS makes it difficult for the GFW to block through fixed domain names or IPs as server addresses frequently change. \\
& Using HTTPS and TLS Encryption\newline
• Comprehensive Encryption: Encrypt DNS queries (e.g., DNS over HTTPS) to prevent DNS request tampering. & 
Encrypting DNS requests prevents the GFW from performing DNS pollution and hijacking, ensuring users receive correct DNS resolutions \cite{1,2,3,4,5}. \\
\midrule
IP Address and IP Range Blocking & 
Dynamic and Random Ports\newline
• Port Randomization: Randomly select ports instead of using common VPN ports to reduce port-based blocking. & 
Random ports make it difficult for the GFW to block VPN traffic through fixed port blocking. \\
& Domain Hiding and Dynamic DNS\newline
• Domain Fronting: Use CDN domain names to hide the actual server addresses. & 
Domain fronting uses large CDN domain names, making it hard for the GFW to identify and block the actual VPN server IP addresses. \\
& Frequent Updates and Protocol Switching\newline
• Dynamic Protocol Switching: Regularly change the protocols and ports used to prevent the GFW from establishing stable blocking patterns. & 
Continuously changing protocols and ports increases the difficulty for the GFW to block, prolonging the effectiveness of bypassing censorship. \\
\midrule
Keyword Filtering and URL Detection & 
Protocol Obfuscation\newline
• Shadowsocks, V2Ray, Trojan\cite{shadowsocks2024, v2ray2024, trojan2024}: Use various encryption and protocols to make traffic difficult to identify as sensitive content. & 
These protocols encrypt and obfuscate traffic content to prevent the GFW from detecting sensitive content through keyword filtering. \\
& Deep Packet Inspection (DPI)\newline
• DPI Countermeasures: Use traffic that mimics common protocols (e.g., HTTPS) to reduce the likelihood of being detected by keyword filtering and DPI. & 
By making VPN traffic appear like regular HTTPS traffic, the risk of being detected by keyword filtering and DPI is lowered. \\
\midrule
Deep Packet Inspection (DPI) & 
DPI Countermeasures\newline
• Obfsproxy, Stunnel: Obfuscate VPN traffic as random data or common protocol traffic, such as HTTPS, to evade DPI detection. & 
These tools alter the characteristics of the traffic, making it difficult for DPI technologies to recognize and classify it. \\
& Using HTTPS and TLS Encryption\newline
• TLS 1.3: Adopt the latest TLS versions to enhance encryption strength and concealment, reducing the likelihood of DPI recognition. & 
Strong encryption makes it difficult for DPI to parse traffic content, thereby making detection and blocking challenging. \\
& Anti-Censorship Protocols\newline
• WireGuard with Obfuscation: Add obfuscation layers to WireGuard traffic, making it difficult for DPI to recognize. & 
Combining WireGuard's efficiency with obfuscation techniques enhances resistance against DPI detection.\cite{wireguard2024} \\
\midrule
Active Probing and Interference & 
Protocol Obfuscation and Anti-Censorship Protocols\newline
• V2Ray, Trojan: Highly customizable protocols that are difficult for the GFW to identify through active probing. & 
These protocols are designed to counter active probing, increasing the difficulty of being identified and blocked by the GFW. \\
& Frequent Updates and Protocol Switching\newline
• Dynamic Protocol Switching: Continuously change protocols and ports, making it difficult for the GFW's active probing to remain effective. & 
Dynamically changing protocols and ports makes it hard for the GFW to establish effective active probing and blocking strategies.\cite{7314859} \\
\bottomrule
\end{tabular}
\end{table*}

\section{Case Analysis of Real-World Strategic Interactions: GFW vs. Tor}

\subsection{Introduction}

The Great Firewall of China (GFW) represents a comprehensive system of internet censorship, combining advanced technical measures with legal and administrative controls. It has evolved significantly since its inception, posing continuous challenges to circumvention tools like Tor. Tor (The Onion Router)\cite{tor2024} is designed to facilitate anonymity and bypass censorship by routing user traffic through a distributed network of volunteer-operated nodes. The ongoing interaction between the GFW and Tor exemplifies the broader conflict between state-controlled internet censorship and tools that champion free access to information\cite{britannicaGFW, wu2023gfw}.

\begin{figure}[h]
    \centering
    \includegraphics[width=\linewidth]{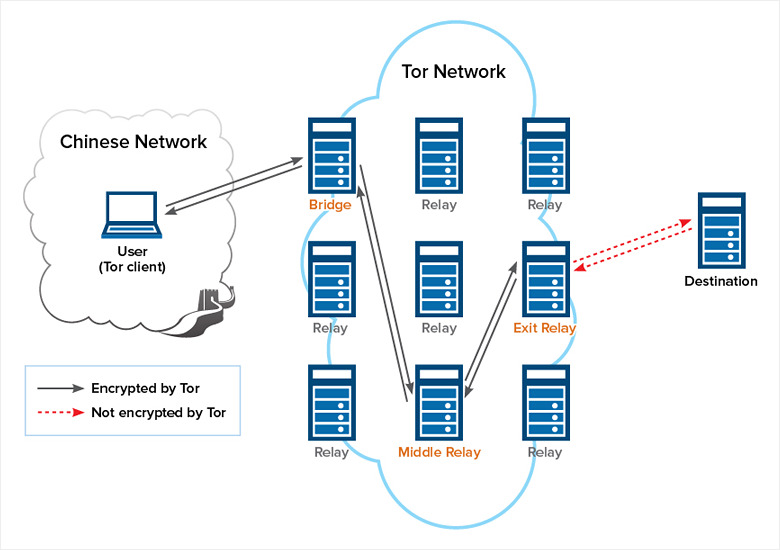}
    \caption{Tor evades censorship using encryption and also preserves anonymity by bouncing requests off several anonymous servers around the world.}
    \label{fig:tor-network}
\end{figure}

\subsection{Technical Strategies of the GFW}

Before 2011, the GFW relied primarily on basic techniques such as keyword filtering and IP blacklisting. However, the introduction of more advanced strategies marked a turning point in its effectiveness. The GFW employs multiple techniques, including Deep Packet Inspection (DPI), active probing, and rapid detection mechanisms, to identify and block Tor traffic. DPI allows the GFW to analyze both headers and payloads of data packets, identifying traffic patterns specific to Tor protocols. Active probing further enables the GFW to block Tor nodes by sending probes and analyzing responses. Since 2011, these advancements have allowed the GFW to detect and block private Tor bridges within minutes of deployment \cite{thousandeyes2024,4476538}.

\begin{figure}[h]
    \centering
    \includegraphics[width=\linewidth]{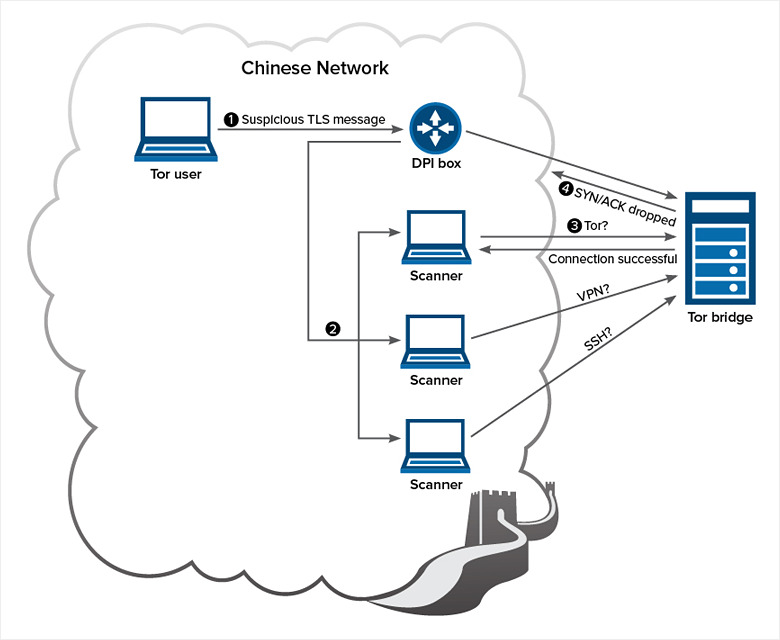}
    \caption{DPI boxes tip off scanners that attempt to connect to the suspected server with a number of different protocols. If a connection attempt succeeds, the proxy is blocked.}
    \label{fig:dpi-scan}
\end{figure}

Figure~\ref{fig:dpi-scan} illustrates how the GFW uses DPI boxes to initiate scanners that systematically probe suspected Tor servers. By sending specific connection attempts across multiple protocols (e.g., VPN, SSH), the GFW can determine the nature of the service and block access accordingly. This process highlights the systematic and multi-layered approach of the GFW in identifying and disrupting Tor operations.

\subsection{Tor’s Countermeasures}

In response to the GFW’s increasingly sophisticated detection mechanisms, the Tor Project and its community have continually refined their suite of countermeasures. As illustrated in Figure~\ref{fig:tor-network}, a central strategy has long been the use of unpublished (bridge) relays—entry nodes to the Tor network not publicly listed in the main Tor consensus. The logic behind bridges is that, by remaining unknown to automated censorship systems, they are less susceptible to large-scale IP blocking. However, studies have shown that as early as 2011, the GFW adapted by pairing Deep Packet Inspection (DPI) with active scanning to detect these covert entry points, a tactic that drastically reduces the anonymity and usefulness of unpublished bridges over time.

When a user in China initiates a connection to what appears to be an unpublished Tor bridge, DPI modules within the GFW identify Tor-like signatures. Immediately following detection, active scanners—distributed across numerous IP addresses within China—attempt to confirm whether the suspected server is indeed running Tor. If the scanner receives a responsive handshake consistent with Tor, the GFW quickly blacklists the bridge’s IP address, often within minutes. Notably, recent measurements indicate that such IP-level blocks can persist for roughly 12 hours, after which the GFW periodically re-probes to determine if the server is still offering Tor connections. This approach significantly constrains the long-term viability of unpublished bridges, as even a single detection can temporarily render a bridge inaccessible to Chinese users.

To combat these evolving tactics, Tor developers have introduced pluggable transports—modular systems like Obfs4, ScrambleSuit, and Meek—that reshape or conceal Tor traffic patterns. Obfs4 randomizes traffic characteristics to evade simple DPI-based fingerprinting, while Meek employs domain fronting to route Tor traffic through reputable CDNs, forcing the GFW to risk collateral damage if it attempts to block it. Although these transports remain effective, their long-term sustainability has been challenged by the GFW’s pressure on CDNs and ongoing improvements in machine learning-based detection \cite{hongkongfp2017}.

Moreover, recent findings suggest that server-side strategies can further frustrate active scanning attempts. For instance, a bridge operator can selectively drop packets from known scanner signatures—distinguishable by certain TCP packet options and MSS values—instead of responding in a manner consistent with Tor. By refusing to acknowledge scanner probes, the bridge avoids triggering long-duration IP blacklisting, effectively leveraging the GFW’s resource-sensitive scanning infrastructure against itself. Such techniques, combined with traditional pluggable transports, represent another layer of defense, enabling Tor bridges to remain operational under the GFW’s stringent surveillance regime.

In summary, the interplay between the GFW and Tor’s countermeasures has evolved from basic IP hiding to a complex arms race involving DPI, active probing, and adaptive transport obfuscation. While unpublished bridges, pluggable transports, and domain fronting have each offered periods of reprieve, the GFW’s active scanning capabilities and IP-based blocking strategies demand continual refinement of circumvention techniques. As research \cite{dunna2018} indicates, practical mitigations—such as traffic dropping and improved distribution of unpublished addresses—are critical components of Tor’s evolving toolkit in resisting state-level censorship.
\cite{tor_blocked_china}

\subsection{Evolution of the Interaction}

The iterative nature of the technological competition between the GFW and Tor highlights the dynamic and high-stakes conflict between censorship tools and anti-censorship efforts. This section outlines key phases of this evolution.

\subsubsection{Pre-2011 Era}

During this period, the GFW relied heavily on static IP blacklisting. Tor bridges provided an effective means of circumvention, requiring minimal adaptation to bypass censorship. The simplicity of the GFW’s methods at the time allowed Tor to maintain consistent access for its users.

\subsubsection{Post-2011 Advancements}

Starting in 2011, the GFW deployed DPI and active probing, significantly reducing the effectiveness of traditional Tor bridges. By analyzing data packets for protocol-specific patterns and probing suspected bridges, the GFW was able to detect and block private bridges rapidly \cite{thousandeyes2024}. These developments forced Tor to adopt more sophisticated countermeasures.

\subsubsection{Domain Fronting and Counteractions}

To address the limitations imposed by the GFW’s advancements, Tor introduced the Meek protocol, leveraging domain fronting to disguise traffic as standard HTTPS connections. This technique routed traffic through popular CDNs, such as Google or Microsoft. However, by 2018, the GFW successfully pressured these CDNs to disable domain fronting, rendering Meek ineffective in highly censored regions\cite{torproject_meek}.

\subsubsection{Machine Learning Integration}

Recent years have seen the GFW incorporate machine learning to enhance its detection capabilities. By analyzing complex traffic patterns in real-time, machine learning models can identify obfuscated Tor traffic with increased precision. This advancement poses new challenges for Tor, necessitating further innovation to counteract these adaptive measures.\cite{hongkongfp2017, Bagui02042017}

\subsubsection{Conclusion of the Interaction}

The ongoing interaction between the GFW and Tor underscores the broader struggle for internet freedom. As both sides continue to innovate, the arms race between censorship tools and anti-censorship technologies exemplifies the high stakes involved in preserving privacy and access to information.

\subsection{Impact and Broader Implications}

The conflict between the GFW and Tor has significant global implications. Techniques pioneered by the GFW, such as DPI and active probing, have been adopted by other nations to implement their censorship systems. Conversely, innovations developed by Tor have advanced privacy tools, benefiting users worldwide. However, this conflict also raises ethical questions regarding the balance between internet freedom and the potential misuse of anonymity tools. Both the GFW’s surveillance practices and the misuse of Tor highlight the complexities of this technological arms race.

\section{Challenges and Bottlenecks in the Obfuscation Arms Race}

\subsection{Limitations of Network Detection}

Detecting VPN traffic at the scale and sophistication of the Great Firewall (GFW) presents substantial technical and economic challenges. While DPI and machine learning-based traffic analysis have advanced considerably, applying these techniques across the GFW’s vast user base and diverse traffic patterns imposes enormous computational overhead. For instance, distinguishing obfuscated VPN traffic from regular encrypted HTTPS sessions requires continuous updates to detection heuristics. This complexity stems not only from the sheer volume of traffic but also from the need to accurately identify increasingly subtle fingerprinting signals.

Moreover, the GFW faces a delicate balance in its detection strategies. Aggressive filtering that mistakenly classifies legitimate encrypted communications as VPN traffic leads to false positives. Such overblocking can disrupt e-commerce transactions, banking services, and secure corporate communications, ultimately diminishing trust in domestic internet infrastructure. Thus, the GFW must manage the trade-off between refining its detection algorithms to minimize circumvention tools and maintaining a low false positive rate that preserves normal network functionality.

\subsection{Bottlenecks in VPN Obfuscation}

From the perspective of VPN providers, advancing obfuscation techniques to outpace GFW detection brings its own set of hurdles. While services such as NordVPN, Surfshark, and ExpressVPN have adopted protocol masking, domain fronting, and randomized traffic patterns, these strategies can degrade performance and reliability. For example, sophisticated obfuscation designed to evade the GFW’s DPI might induce higher latency or packet loss, impairing user experience within China and making the service less appealing for everyday activities that require stable, low-latency connections.

In addition, not all environments tolerate these obfuscation methods gracefully. Corporate firewalls or legacy systems may conflict with obfuscation layers designed for the Chinese censorship context, reducing compatibility in other scenarios. This fragmentation means VPN providers must continually refine their approaches, seeking the delicate equilibrium between robust anti-censorship capabilities and consistent, user-friendly performance\cite{285411, Ramesh2022VPNalyzer, cloudflare2021vpn}.

\begin{figure}[ht]
\centering
\includegraphics[width=8.5cm]{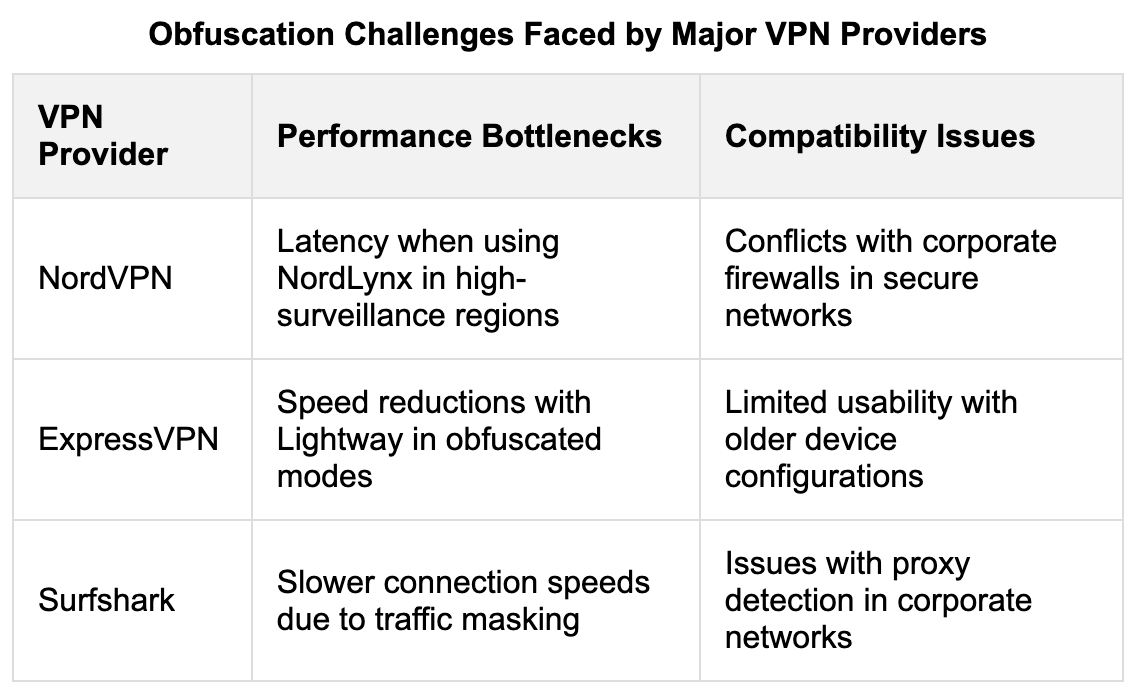}
\caption{Obfuscation Challenges Faced by Major VPN Providers in Circumventing the GFW}
\label{fig:vpn_obfuscation_challenges}
\end{figure}

These challenges underscore the need for strategic innovation in VPN design, where modular, easily updatable obfuscation layers can adapt quickly to the GFW’s evolving detection techniques.

\subsection{Privacy and Ethical Considerations}

The escalating arms race between VPN providers and the GFW’s network detection apparatus is not purely technical—it carries profound privacy and ethical implications. On one side, VPNs champion user autonomy, allowing individuals to bypass state-imposed restrictions and access information freely. In environments like China, where the GFW enforces stringent content controls, VPNs serve as a lifeline for journalists, researchers, and citizens seeking unbiased news and external perspectives. Yet, the very encryption and obfuscation that protect user anonymity and freedom also facilitate illicit activities. Bad actors might exploit these same tools for criminal conduct or to evade legitimate law enforcement, complicating the ethical assessment of VPN usage.

Conversely, the GFW’s increasingly intrusive detection measures raise concerns about overreach and disproportionate surveillance. Overly broad monitoring schemes risk infringing on users’ fundamental rights to privacy and freedom of expression, and may lead to collateral damage affecting ordinary citizens and businesses. Navigating these ethical dilemmas calls for nuanced policies and international dialogue, aiming to maintain security and stability without sacrificing the digital liberties that VPNs—despite their complications—ultimately strive to uphold.

\section{Conclusion}
The evolution of China’s Great Firewall (GFW) encapsulates an ongoing and deeply complex struggle at the intersection of technology, policy, and human rights. Over time, the GFW has expanded beyond simple IP blacklisting and keyword filtering into a multifaceted censorship apparatus that leverages DNS pollution, IP and URL filtering, Deep Packet Inspection, active probing, and increasingly machine learning–based detection methods. This continuous refinement reflects not only the state’s determination to control information flows but also the adaptability and ingenuity of its censorship infrastructure.

In parallel, the community of developers, researchers, and users dedicated to circumventing these barriers has similarly advanced their techniques. Early approaches to obfuscating VPN traffic have given way to more sophisticated mechanisms, from protocol mimicry and dynamic port switching to domain fronting and emerging anti-censorship protocols. These tactics have, at times, successfully restored a degree of online anonymity, ensuring that knowledge, communication, and collaboration can traverse national and ideological boundaries.

Yet, the power dynamic remains fundamentally asymmetric. Armed with vast computational resources, strategic partnerships with major technology firms, and legal authority to enforce and refine its censorship, the GFW wields a potent advantage. By contrast, circumvention communities rely on open collaboration, rapid iteration, and decentralized development—qualities that foster adaptability but can lack the scale, coherence, and resources enjoyed by state-driven censorship efforts. Over time, both sides will likely continue to refine their approaches, marking this contest as a perpetual arms race rather than a finite battle.

As global internet governance debates intensify, the lessons drawn from the GFW’s evolution hold relevance far beyond China’s borders. The engineering insights, user behaviors, legal frameworks, and ethical dilemmas that have emerged from this environment inform broader discussions about the future of digital spaces. Ensuring that global networks remain accessible and free from unwarranted intrusion will require sustained collaboration across governments, academia, industry, and civil society. Continued innovation and knowledge exchange are essential for building resilience against censorship and preserving the principle of an open internet—one in which information and ideas can circulate freely, even in the face of the most determined efforts to constrain them.

\vspace{12pt}

\end{document}